\documentclass[manuscript,screen]{acmart}
\pdfoutput=1
\usepackage{booktabs}

\usepackage{subcaption}
\usepackage{tabularx}
\usepackage{multirow}
\usepackage[T1]{fontenc}
\usepackage{graphicx}
\usepackage[title]{appendix}
\begin{document}

\title{\underline{\textit{C}}ollaborative \underline{\textit{H}}uman-\underline{\textit{AI}} \underline{\textit{R}}isk \underline{\textit{A}}nnotation: Co-Annotating  Online Incivility with \textit{CHAIRA}}


\author{Jinkyung Katie Park}
\affiliation{%
  \institution{Clemson University}
  \city{Clemson}
  \country{USA}}
\email{jinkyup@clemson.edu}

\author{Rahul Dev Ellezhuthil}
\affiliation{%
  \institution{Rutgers University}
  \city{New Brunswick}
  \country{USA}}
\email{re263@rutgers.edu}

\author{Pamela Wisniewski}
\affiliation{%
  \institution{Vanderbilt University}
  \city{Nashville}
  \country{USA}}
\email{pamela.wisniewski@vanderbilt.edu}

\author{Vivek Singh}
\affiliation{%
  \institution{Rutgers University}
  \city{New Brunswick}
  \country{USA}}
\email{v.singh@rutgers.edu}


\begin{abstract}
Collaborative human-AI annotation is a promising approach for various tasks with large-scale and complex data. Tools and methods to support effective human-AI collaboration for data annotation are an important direction for research. In this paper, we present \textit{CHAIRA}: a \textit{C}ollaborative \textit{H}uman-\textit{A}I \textit{R}isk \textit{A}nnotation tool that enables human and AI agents to collaboratively annotate online incivility. We leveraged Large Language Models (LLMs) to facilitate the interaction between human and AI annotators and examine four different prompting strategies. The developed CHAIRA system combines multiple prompting approaches with human-AI collaboration for online incivility data annotation. We evaluated CHAIRA on 457 user comments with ground truth labels based on the inter-rater agreement between human and AI coders. We found that the most collaborative prompt supported a high level of agreement between a human agent and AI, comparable to that of two human coders. While the AI missed some implicit incivility that human coders easily identified, it also spotted politically nuanced incivility that human coders overlooked. Our study reveals the benefits and challenges of using AI agents for incivility annotation and provides design implications and best practices for human-AI collaboration in subjective data annotation.
\end{abstract}
\maketitle





\section{Introduction}


Online incivility refers to features of discussion that convey an unnecessarily disrespectful tone toward the discussion forum, its participants, or the topic \cite{coe2014online}. 
Considering adverse effects~\cite{han2018civility,gervais2015incivility}, it is important to identify and moderate incivil comments on social media platforms as well as to understand the nature and characteristics of incivility \cite{matias2019preventing,jhaver2018online,coe2014online,oz2018twitter}, both of which require the annotation of digital trace data.
This annotation process often includes crowdsourced workers ~\cite{hosseinmardi2015analyzing}, a team of researchers and research assistants \cite{singh2017toward,park2023towards}, and/or domain experts ~\cite{park2023misinformation}. 
It involves an intensive and collaborative process of training, consensus-building, and quality control among multiple coders; therefore, it can be costly and time-consuming, while still yielding uneven levels of inter-coder agreement ~\cite{coe2014online,rains2017incivility}. There is a growing need for innovative and efficient methods to support human coders in annotating large corpora of online data, which can have a significant methodological impact on information science research.

In this study, we explore the use of Large Language Model (LLM)-based Conversational Agents (CAs) as AI-based co-coders for annotating online incivility data. 
We focus on LLM-based CAs because they have shown promising results in text annotation tasks due to their accuracy and adaptability (e.g.,~\cite{zhang2023qualigpt,huang2023chatgpt,liu2023summary,zhang2022would,amin2023will,kuzman2023chatgpt}). Moreover, LLMs can be adapted through finetuning or prompting for specific domains~\cite{song2024preference}, setting a new standard for what is achievable in natural language tasks. However, there are challenges in the use of LLM-based CAs in annotating textual data for more contextualized constructs~\cite{amin2023will,huang2023chatgpt}, indicating the need for human-AI collaboration on subjective and contextualized annotation tasks.


In this paper, we present \textit{CHAIRA: a Collaborative Human-AI Risk Annotation tool} that enables human and AI agents to co-annotate online incivility. We share early results from the design and implementation of a CHAIRA that we developed to interact with human coders and provide suggestions and explanations for annotating online incivility. 
Using 457 user comments with ground truth labels (e.g., civil vs. uncivil), we experiment with four types of prompting methods with different levels of information exchange between the human coder and CHAIRA. Using 10\% of the data (\textit{n} = 50), we established inter-rater agreement between the human coder and CHAIRA to observe how different types of promoting methods impact data annotation results. We analyzed the conversation log between human coders and CHAIRA to have qualitative insights into how the quality of annotations changes with different prompting approaches. 
As such, we use a mixed methods approach to address the research questions:


\begin{itemize}
\item\textit{\textbf{RQ1}: How do different types of prompting methods influence inter-rater reliability of human-AI collaborative data annotation results? }

\item\textit{\textbf{RQ2}: How do different types of prompting methods influence the quality and rationale for human-AI collaborative data annotation results?}





\end{itemize}

By answering the above research questions, we address the overarching question of: \textit{What is the optimal prompting approach and best practices to make the performance of human-AI collaboration similar to that of human-human collaboration?}
We found that CHAIRA's performance in terms of inter-coder agreement with human coders improved with more detailed prompts. The most advanced model, the Two-stage Few-shot Chain of Thought, nearly matched the agreement levels seen between two human coders reported in previous studies. While the AI agent missed some implicit incivility that human coders easily identified, it also spotted politically nuanced incivility that human coders overlooked.
Our work provides design insights and best practices for human-AI collaboration in subjective data annotation tasks. It introduces a novel system for human-AI collaboration and applies different prompt engineering approaches to optimize incivility annotation. These findings are applicable beyond online incivility scenarios, offering a path for scalable annotation in subjective or low-resource settings. As such, our work contributes to the iConference community by empirically demonstrating the potential of human-AI collaboration in the context of subjective digital trace data annotation. Particularly, we contribute to the iConference community's focus on addressing multifaceted dimensions of AI to foster a deeper understanding of their benefits, challenges, and broader implications.

\section{Related Work}

\subsection{Conversational Agents as Annotators}


Conversational Agents (CAs) are systems enabled with the ability to interact with the users using natural human dialogue ~\cite{rheu2021systematic}. 
After the recent release of various Large Language Model (LLM)-based Conversational Agents (CAs) (e.g., ChatGPT~\cite{OpenAI2023}), research communities are increasingly experimenting with data annotation tasks such as annotating political stance and sentiment of textual data~\cite{liu2023summary,zhang2022would,amin2023will,kuzman2023chatgpt}. Emerging literature suggests that LLM-based CAs can be useful for text classification tasks (e.g.,~\cite{zhang2023qualigpt,huang2023chatgpt,liu2023summary,zhang2022would,amin2023will,kuzman2023chatgpt}). For instance, Zhang et al.~\cite{zhang2022would} show that ChatGPT was able to annotate the political stance of the tweets with an average accuracy above 70. Moreover, LLMs can be adapted through finetuning or prompting for specific domains \cite{song2024preference}, setting a new standard and expectations for what is achievable in natural language tasks. With proper fine-tuning, LLMs are known to even outperform crowdsourced annotators \cite{gilardi2023chatgpt}. As such, advances in LLMs such as GPT-4 showed a promising opportunity for data annotation at scale due to their ability to automate annotation tasks \cite{zhang2022automatic}. However, there are challenges in the use of LLM-based CAs in annotating textual data for more contextualized constructs. For instance, Amin et al.~\cite{amin2023will} showed that ChatGPT's accuracy for subjective tasks such as the five personality and suicide ideation classifications was lower than the baseline machine learning methods. As such, early empirical research demonstrated the limitations of subjective and contextualized annotation tasks entirely, indicating the need for human-AI collaboration on such tasks.


\subsection{Human-AI Collaboration on Annotation}

As LLMs-based conversational agents have shown the ability to interact with humans and work with examples in various domains~\cite{mackeprang2019discovering,lai2022human,tang2024pdfchatannotator,kim2022stylette}, researchers are exploring the potential of human-AI collaboration on various tasks such as online content moderation~\cite{lai2022human}, thematic analysis of qualitative data~\cite{zhang2023qualigpt,jiang2021supporting}, disease preveition~\cite{lu2024differences}, and crowdsourcing~\cite{tamura2024influence}. 
For instance, Lai et al. developed a collaborative model in which humans and AI work together to decide on trustworthy regions of AI before deployment. They showed that the human-AI team model outperformed human-only or AI-only models in content moderation tasks. Similarly, Zhang et al. explored the potential of LLM-based CAs as collaborative tools for qualitative data analysis and highlighted the efficiency of reducing time and labor for such analysis~\cite{zhang2023qualigpt}. 
Yet, their performance in collaborative co-annotation exercises for online risk where different facets of co-annotation are important is understudied. This gap is pertinent because co-annotation tasks need to support an interactive discussion to help generate a rationale for the various decisions, particularly in the context of highly contextualized online risk behavior~\cite{clay2003context}
which can entail disagreement even among human coders. Yet, the disagreement must not lead to capitulation, instead, inspire better methods of automated analysis. Therefore, the combination of manual and automated content analysis is suggested as the gold standard for identifying subjective concepts such as online risk~\cite {esau2022content}. 

In this work, we explore the potential of using LLM-based CAs to assist human coders in annotating subjective, nuanced online conversations. We expect that LLM-based CAs can support high-quality annotations with explanations when provided with proper instructions and examples. This approach could enhance scalability and help capture nuances that human coders might miss due to cognitive limits. We experimented with four prompting methods to assess their impact on annotation results and how co-annotation can improve subjective data annotation. By examining agreements and disagreements between human coders and LLM-based CAs, we explored how their strengths and weaknesses can complement each other.

\section{Methods}

\subsection{Design and System Implementation of CHAIRA}
CHAIRA is an online annotation tool that integrates an LLM-based conversational agent to support human-AI collaboration on online risk data annotation. Below, we describe how we designed and implemented CHAIRA in detail. 

\paragraph{\textbf{System Implementation and Dataset}} 
We leveraged the GPT 3.5 Turbo (the underlying model for OpenAI's ChatGPT) as the LLM of choice due to its popularity and ease of use. We developed a custom annotation interface on top of OpenAI's API to support human-AI co-annotation. The interface was developed in React and deployed on AWS. We used S3 buckets to store the dataset and AWS Lambda to evaluate the dataset. 
We used the dataset collected in prior work in which researchers explored the effectiveness of embedding positive background images on online discussion forums in reducing online incivility~\cite{park2022background}. The data comprised 457 comments collected from 105 users who participated in an online experiment and were annotated for online incivility by the two human coders. 
In the prior work~\cite{park2022background}, a researcher designed a codebook to annotate 457 use comments into civil vs. incivil and worked with a human coder (i.e., research assistant) to establish the reliability of the incivility coding scheme. After the researcher and the coder had multiple training sessions, 10\% of the comments (\textit{n} = 45) were coded to establish interrater reliability. The reported interrater reliability scores in prior work were 0.88 (percent agreement) and 0.76 (Cohen's Kappa). Once the interrater reliability was established, the researcher coded the rest of the data. In the final dataset, 55\% were reported to be civil cases (\textit{n} = 250) while 45\% were incivil cases (\textit{n} = 207)~\cite{park2022background}.

In this work, we split the labeled dataset (457 use comments) into training (5\%), validation (10\%), and test (85\%) datasets using a stratified random sampling approach. Consistent with common practices in human-human collaborative coding, less than $5\%$ of data (20 user comments) were allocated to the training dataset to be used as examples and initial instructions (similar to training sessions for human-human coding). Then each prompt was passed to 50 samples (approx. $10\%$ of data) allocated to the validation dataset to evaluate the inter-rater reliability between the human coder and CHAIRA. The rest of the comments (387 comments) were allocated to the test dataset to evaluate human-AI agreement on final online risk data annotation results. 

Following the common human-human coding practices, we split our training, validation, and test datasets to be independent of each other. For instance, samples from either the validation or test split were not mixed with the training split. In addition, only comments from the training set can be added as examples in the prompts to fine-tune CHAIRA. Similarly, samples from the validation set can be used to interact with CHAIRA but cannot be added as examples in a prompt. Samples from the test split cannot be loaded into the prompt. 

\begin{figure}[t]
    \centering
    \includegraphics[width=\textwidth]{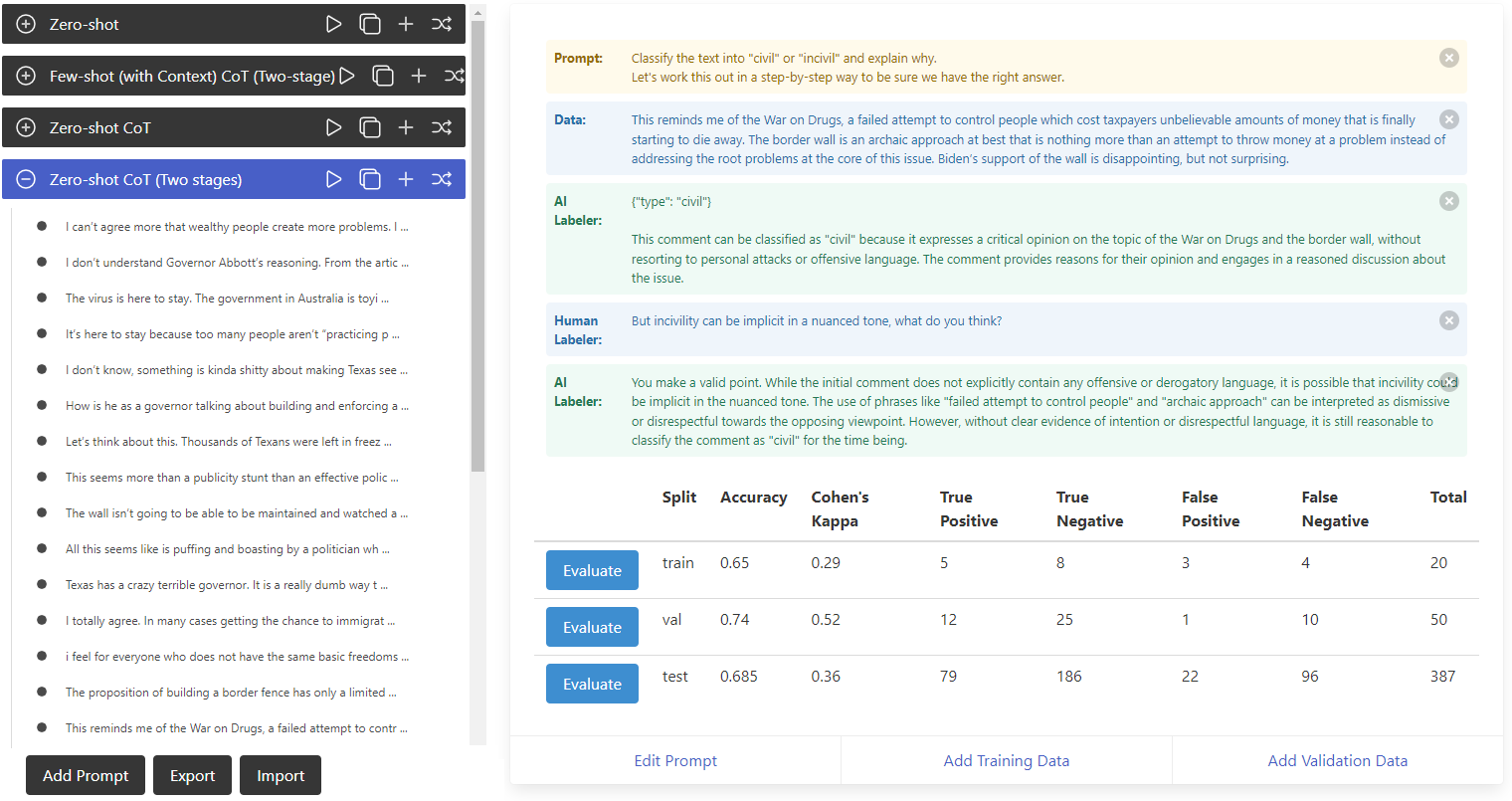}
    \caption{Overview of CHAIRA web interface. A list of designed prompts is shown on the left side, while the prompt, conversation log, and evaluation metrics/results are shown on the right side.}
        \label{fig:UI full}
\end{figure}

\paragraph{\textbf{Design of Web Interface}}
The web interface of CHAIRA (see Figure~\ref{fig:UI full}) provides an overview of the layout for a human coder to interact with CHAIRA and design prompts. 
The left side of the interface shows a list of different prompts created for online risk annotation tasks in black labels. Once the human coder clicks a specific prompt, the label of the prompt turns blue to indicate that it is an active prompt. The right side of the interface shows the chosen prompt, comment data, conversation log between the human coder and AI agent, and inter-rater agreement. Below, we zoom into the major components of the CHAIRA interface to describe how each component was designed to facilitate human-AI collaboration on online risk data annotation tasks. 


\paragraph{Creating Prompts}
\begin{figure}[t]
    \centering
    \includegraphics[width=0.8\textwidth]{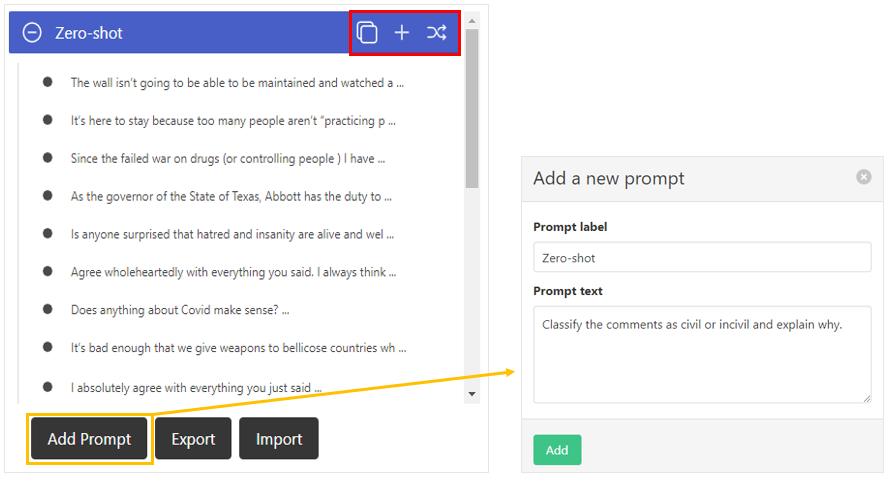}
    \caption{Features to create and manage prompts to interact with CHAIRA}
    \label{fig:prompt}
\end{figure}
On the left side of the interface, the human coders can create new prompts to interact with an AI agent by clicking ``Add Prompt'' (Figure~\ref{fig:prompt}). Once clicking the add prompt, a new text box appears where human coders can add the name of a certain prompt in the ``Prompt label'' box and add the content of the prompt in the ``Prompt text'' box to create a new prompt. 
Once new prompts are created, the human coders can add sample comments to test with the new prompts. A double arrow icon (on the left side of the red box in Figure~\ref{fig:prompt}) helps human coders to randomly sample comments from the training data. The user comments are added within the threads under each prompt with bullet points. Beyond the labeled dataset (457 user comments), the human coders can manually add new comments to label incivility using the same prompt by clicking a plus icon in the middle in the red box in Figure~\ref{fig:prompt}. Yet, these comments are not included when evaluating the prompt for inter-rater agreement with a human coder. 
Human coders can create copies of existing prompts by clicking the double squire icon on the left in the red box in Figure~\ref{fig:prompt}. We chose icons for the above three features as there is limited space allocated in the prompt label (Figure~\ref{fig:UI full}). The ``Export'' feature helps human coders download the prompt and conversation log data in a JSON file format, while the ``Import'' feature helps the opposite, uploading JSON files to the interface.

\begin{figure}[h]
    \centering
    \includegraphics[width=\textwidth, height=2in]{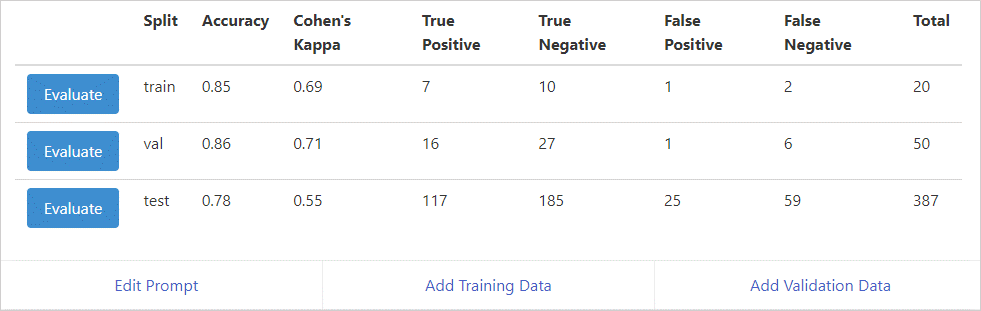}
    \caption{Features to assess inter-rater reliability between the human coders and CHAIRA}
    \label{fig:IRR}
\end{figure}

\paragraph{Evaluating Inter-rater Agreement}
Inter-rater agreement between the human coder and the AI agent can be assessed using the ``Evaluate'' buttons (Figure~\ref{fig:IRR}). 
To report inter-rater agreement between the human coder and CHAIRA, we used percent agreement and Cohen's Kappa, following the practices in the literature on qualitative content analysis~\cite{tinsley2000interrater,mcdonald2019reliability}. 
``Add Training Data'' loads all 20 user comment data from the training dataset to be used as examples and initial instructions. Once the human coders click the button, the interface creates a thread under the prompt to display all 20 user comment data (see the left side of Figure~\ref{fig:UI full}). 
``Add Validation Data'' loads all 50 samples allocated in the validation dataset to establish the inter-rater agreement between the human coders and CHAIRA. 
After looking at the evaluation results, the human coder can edit the prompt by clicking the ``Edit Prompt'' button. The interface can evaluate multiple prompts at the same time, which supports its scalability.

\paragraph{Human-AI Interaction}

\begin{figure}[t]
    \centering
    \includegraphics[width=\textwidth, height=10cm]{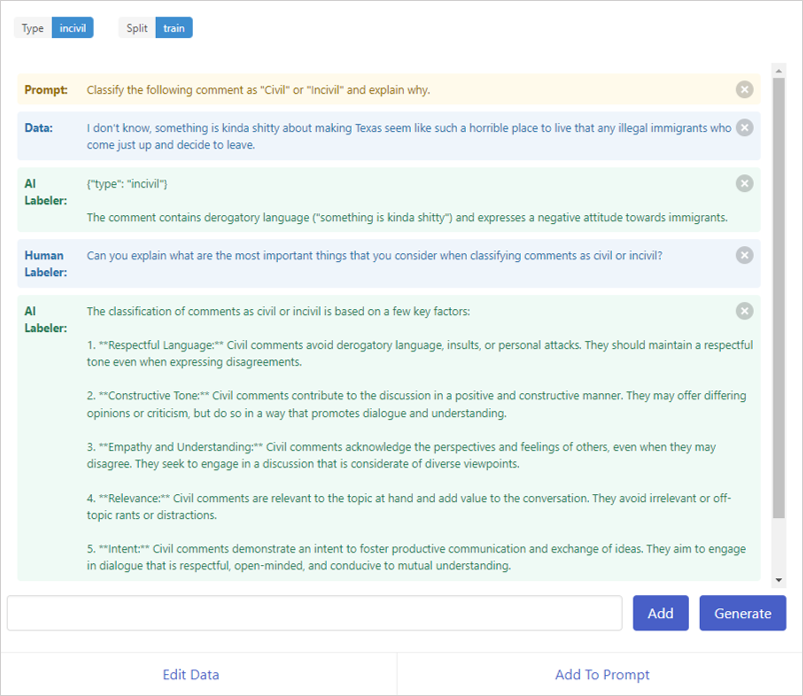}
    \caption{Features to support interactive communication between the human coders and CHAIRA}
    \label{Fig:Interaction}
\end{figure}

Once the interface loads user comment data from the training set, the human coders can have interactive conversations with the AI agent on the right side of the interface. Once the human coder clicks a certain user comment under the prompt, the interface shows the prompt, comment data, and response from CHAIRA (Figure~\ref{Fig:Interaction}).
The ``Type'' icon on the top left indicates an incivility label annotated by the human coder in prior work. The ``Split'' icon indicates to which dataset the user comment data belongs. As seen from Figure~\ref{Fig:Interaction}, ``incivil'' and ``train'' means that the user comment data came from the training dataset and was annotated as incivil by the human coder.
Right below these icons, the text in the prompt is shown as brown text with the label ``Prompt'' on a yellow background. User comment data to annotate for incivility comes next as blue text with the label ``Data'' in a light-blue background. Then the incivility labels and the rationales for the decision generated by CHAIRA follow as green text with the label ``AI Labeler'' in a light-green background. 
The three components above are automatically generated when evaluating the inter-rater reliability for each prompt. 

After reviewing the initial response generated by CHAIRA, the human coders can start a conversation by adding queries in the textbox and clicking the ``Add'' button next to the textbox. Then, when the human coders click the ``Generate'' button, CHAIRA generates responses to the given queries. The queries asked by the human coders appear with the label ``Human Labeler'' in a light-blue background, while the answers generated by CHAIRA appear with the label ``AI Labeler'' in a light-green background (Figure~\ref{Fig:Interaction}). We designed text generated by CHAIRA to appear as green text in light-green backgrounds while text submitted by the human coders to appear as blue text in light-blue backgrounds to help human coders distinguish text generated by the two parties.

Through these interactive conversations, additional instructions and examples are exchanged between the two. Once the human coder decides that a reasonable agreement was achieved, a conversation log between the human coder and CHAIRA can be added as a prompt by clicking ``Add To Prompt'' (Two-stage prompt in the next section).  Following the common practices in human-human collaborative coding in which approximately 5-10\% of the data is used for training and consensus building, we designed only conversation around 20 user comments in the training dataset to be added as prompts using the ``Add To Prompt'' feature. Finally, the ``Edit Data'' feature helps the human coders edit the user comment data.



\subsection{Prompt Engineering Approaches}

We conducted experiments with four different prompting engineering approaches: zero-shot, definition, few-shot, and two-stage few-shot Chain-of-Thought (CoT). We designed four different prompting approaches with varying levels of interaction between the human coder and the LLM-based agent. 
We used the same coding scheme as applied in the previous literature~\cite{park2022background} to design the prompts. In our prompts, incivility was defined as ``the feature of discussion that conveys an unnecessarily disrespectful tone toward the discussion forum, its participants, or its topic''~\cite{coe2014online}, with six different categories: name-calling, aspersion, lying, vulgarity, pejorative for speech, and others (Table~\ref{tab:definition}). 
The reported inter-rater agreement between the two human coders was 0.88 (percent agreement) and 0.76 (Cohen's Kappa)~\cite{park2022background}.
\begin{table}[h]\small
\caption{Definition and Examples of Types of Incivility (Park et al.~\cite{park2022background})}
\label{tab:definition}
\begin{tabular}{p{1.7cm}|p{4.9cm}|p{5.2cm}}
\hline
\textbf{Category} & \textbf{Description} & \textbf{Example} \\ \hline
\textbf{Name-calling} & Mean-spirited or disparaging words directed at a person or group of people. & “At least the morons in the state capital no longer have control of this process!” \\ \hline
\textbf{Aspersion} & Mean-spirited or disparaging words directed at an idea, plan, policy, or behavior. An aspersion may be both explicit and implicit. & “It beckons the memories of Trump’s silly border wall and the incredible waste of resources that was.” \\ \hline
\textbf{Lying} & Stating or implying that an idea, plan, policy, or public figure was disingenuous. & “Government is wrong, is corrupt, is lying, is deceiving the people, and is violating our constitution” \\ \hline
\textbf{Vulgarity} & Using profanity of language that would not be considered proper in professional discourse. & “Am I possibly the only person here who thinks this shit is funny as hell?” \\ \hline
\textbf{Pejorative for speech} & Disparaging remark about the way in which a person communicates. & “Quit crying over the spilled milk of.” \\\hline
\textbf{Others} & All comments that may be deemed incivil, but do not fall into any of the previous categories of incivility. & “Hahahahahahahahahahaha,, really crack me open this one.” \\ \hline
\end{tabular}
\end{table}

\paragraph{\textbf{Zero-shot prompting}} 
In zero-shot prompting, the model is only given a simple instruction describing the task. This method is considered convenient and has the potential for robustness~\cite{brown2020language}. The instruction used in the zero-shot prompt is as follows: \textit{``Classify the text into ``civil'' or ``incivil'' and explain why.''} 

\paragraph{\textbf{Definition prompting}} 
In Definition prompting, along with the instruction of classifying a comment as ``civil'' or ``incivil,'' we provided the definition of incivility and brief descriptions of six categories of incivility (see Table~\ref{tab:definition}). 

\paragraph{\textbf{Few-shot prompting}}
With few-shot prompting, models are given a few demonstrations of the task~\cite{brown2020language}, in our case, examples of incivility. With this approach, we provided the model with the definition of incivility, descriptions of the six categories of incivility, examples of the six categories of incivility (Table \ref{tab:definition}), and the instructions for the task.

\paragraph{\textbf{Two-stage Few-shot Chain of Thought}}
Finally, we used a two-stage few-shot Chain-of-Thought (CoT), a few-shot-based prompting for the chain of thought reasoning. Chain-of-thought (CoT) prompting~\cite{wei2022chain} modifies the answers in few-shot examples to step-by-step answers by adding an instruction such as ``Let's think step by step'' to the original prompt to elicit reasoning in LLMs~\cite{wei2022chain,kojima2022large}.
The first prompt is the reasoning extraction where we used CoT to elicit reasoning from LLM. The second prompt consists of the first prompt and the answers generated from the first prompt. 

\begin{figure}[t]
    \includegraphics[width=\textwidth]{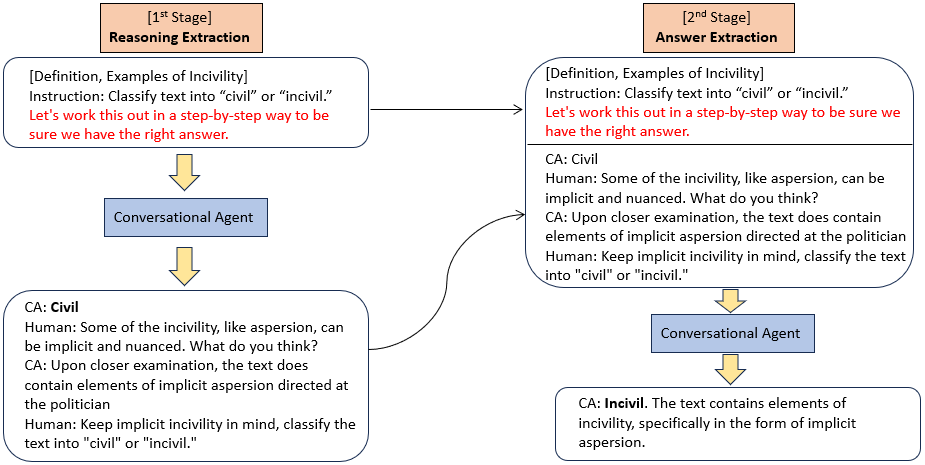}
    \caption{Pipeline of Two-stage Chain of Thought prompting. Human feedback on Phase 1 and CA responses are prepended to the input for Phase 2 sent to CA.}
    \label{fig:pipleline}
\end{figure}

Figure \ref{fig:pipleline} shows a visualization of the two-stage few-shot CoT approach applied.  In the first prompt, we provided the same instructions as in few-shot prompting and added the final line that says, ``Let's work this out in a step-by-step way to be sure we have the right answer'' suggested by Zhou et al.~\cite{zhou2022large}. 
Next, we reviewed the responses generated by CHAIRA from the first prompt and did the error analysis. We chose one example of false positive cases (i.e., CHAIRA output = civil, human ground truth = incivil) and pointed the model to recognize what it is missing in its answers (i.e., implicit aspersion). Once CHAIRA  generated the responses that matched with human responses, we added the conversation log (Table~\ref{tab:conversation log} in Appendix A) to the prompt. 
A summary of the four prompting approaches we used in this case study is present in Table~\ref{tab:prompts}. 

\section{Results}
Our methodology, system implementation, and prompting strategies showed that practical systems for human-AI collaboration in online risk annotation are feasible. To investigate RQ1, we measured the inter-coder agreement between the human coder and the AI agent. The agreement increased with the amount of detail in the prompts. The inter-rater agreement increased with additional information given in the prompts. Two-stage CoT yielded the highest performance (Cohen's Kappa = 0.71), yet, lower than the levels observed for the baseline, two human coders (Cohen's Kappa = 0.76)  ~\cite{park2022background} (see Table~\ref{tab:prompts}). 

\begin{table}[h]\small
\caption{Summary and Performance of the Four Prompting Approaches}
\begin{tabular}{p{1.6cm}p{1.3cm}p{1.3cm}p{1.3cm}p{1.3cm}|p{1.8cm}p{1.8cm}}
\cline{2-7}
              & \multicolumn{4}{c|}{Prompt}                                                                                                 & \multicolumn{2}{c}{Performance}                                          \\ \cline{2-7} 
              & \multicolumn{1}{c}{Instruction} & \multicolumn{1}{l}{Definition} & \multicolumn{1}{l}{Example} & \multicolumn{1}{p{1.9cm}|}{Conversation Log} & \multicolumn{1}{p{1.8cm}}{Percent Agreement} & \multicolumn{1}{p{1.8cm}}{Cohen's Kappa} \\ \hline
Zero-shot     & \multicolumn{1}{c}{X}                               &                                &                             &                                       & \multicolumn{1}{c}{0.66}                                 & \multicolumn{1}{c}{0.26}                              \\
Definition    & \multicolumn{1}{c}{X}                               & \multicolumn{1}{c}{X}                              &                             &                                       & \multicolumn{1}{c}{0.72}                                 & \multicolumn{1}{c}{0.48}                              \\
Few-shot      & \multicolumn{1}{c}{X}                               & \multicolumn{1}{c}{X}                              & \multicolumn{1}{c}{X}                           &                                       & \multicolumn{1}{c}{0.78}                                 & \multicolumn{1}{c}{0.54}                              \\
Two-stage CoT & \multicolumn{1}{c}{X}                               & \multicolumn{1}{c}{X}                              & \multicolumn{1}{c}{X}                           & \multicolumn{1}{c|}{X}                                     & \multicolumn{1}{c}{\textbf{0.86}}                                 & \multicolumn{1}{c}{\textbf{0.71}}                              \\ \hline
Baseline & \multicolumn{1}{c}{X}                               & \multicolumn{1}{c}{X}                              & \multicolumn{1}{c}{X}                           & \multicolumn{1}{c|}{X}                                     & \multicolumn{1}{c}{0.88}                                 & \multicolumn{1}{c}{0.76}                              \\ \bottomrule
\end{tabular}
\label{tab:prompts}
\end{table}

To further understand the impact of different prompting methods on the annotation quality and rationale (RQ2), we discuss some of the common themes observed in the logs of interaction between the human coder and CHAIRA. 

\subsection{CHAIRA performed better with more human-AI interaction in the prompts, particularly for annotating implicit incivility}

Overall, CHAIRA did a better job at annotating explicit incivility and explaining the reasoning behind them with more information provided in the prompts. 
For instance, even with the zero-shot prompting approach, CHAIRA pointed to the exact incivil expressions and recognized that such expressions reflect a negative attitude towards immigrants. In addition, without any information about incivility, CHAIRA was able to note the use of more nuanced incivility, such as sarcasm, when nuanced expressions are combined with explicit personal attacks and insults (examples in Table~\ref{tab:uncivil example} in Appendix).  

At the same time, we observed some common issues throughout all the prompting approaches: CHAIRA did not recognize implicit and nuanced incivil expressions in the texts, even with the information  ``an aspersion may be both explicit and implicit'' given in the prompt.
Therefore, when designing the two-stage few-shot CoT prompt, we reminded CHAIRA that aspersion can be implicit and nuanced through interactive conversation. 
Only after the interactive conversation between the human coder and CHAIRA about implicit aspersion was added to the prompt (i.e., the two-stage few-shot prompting approach), CHAIRA started to distinguish implicit and nuanced yet incivil expressions (examples in Table~\ref{Example responses} in Appendix), which were frequent in our dataset.


\subsection{The output label remained the same but the reasons changed with different prompts}
We observed that while the output label remained the same the reasons changed with more information in the prompts.
For instance,
with the zero-shot prompt, CHAIRA mainly focused on the use of language and tone of the text, whereas with the definition prompt, CHAIRA considered whether the text fell under any of the six categories of incivility (Table~\ref{tab:uncivil example}). The rationales provided were similar for definition and few-shot prompting approaches.  
A similar trend was observed for incivil cases, where CHAIRA provided specific reasons by pointing to the specific category (i.e., lying) and the context of why it falls under the category with the definition and few-shot prompting approaches, while CHAIRA provided more generic reasons (i.e., use of personal attacks or offensive language) with the zero-shot approach.

Overall, CHAIRA could provide human coders with helpful context on US politics, immigration policy debates, and incivil expressions that may be overlooked. For example, the text \textit{“And around Nancy’s wall on Capitol Hill. Make that wall to keep them in”} was annotated as civil by a human coder, as it did not contain explicit or implicit incivility.
However, with zero-shot and definition prompts, CHAIRA recognized “them” as immigrants, and the comment about building a wall to “keep them in” was interpreted as disrespectful. Using a two-stage few-shot CoT approach, CHAIRA’s rationale became even more detailed: “The comment mocks Nancy Pelosi and implies hostility toward immigrants.”
The human coder missed that “Nancy” referred to a political figure and failed to recognize the disparaging tone. Although another coder might have annotated it differently, the above example shows that human coders have limits in awareness and cognitive capacity and AI can complement such limitations. As such, with the highest level of human-AI interaction (two-stage CoT), CHAIRA effectively discerned both the tone and the target of incivility, providing crucial context in annotating political incivility.

\subsection{Yet, sometimes, CHAIRA did not fully understand the information in the prompts and text to annotate.}
Unlike human conders, we observed some cases where CHAIRA did not accurately catch the textual information given in the prompts. For instance, CHAIRA lacked an understanding of the description and example of the ``Pejorative for Speech'' category. We found some of the responses that CHAIRA mistakenly understood the pejorative for speech as the use of a sarcastic tone in the text, as opposed to its actual concept, a disparaging remark about how a person communicates.
Similarly, CHAIRA sometimes lacked an understanding of the given text compared to human coders, particularly for short texts. For instance, 
with the definition prompting approach, CHAIRA struggled to annotate a short text such as ``What is your solution?'' hence, annotated it as ``unclear.'' 
With the few-shot prompt, CHAIRA annotated the same text as incivil because the text contains examples of name-calling and aspersions. 
However, the incivil expressions referred to in this response were from the examples given in the prompt (i.e., instructions), not from the given text to annotate. Overall, sometimes CHAIRA could not understand the information in the prompts or texts to annotate and hence, was unable to annotate the text appropriately.

\section{Implications for Building Human-AI Collaborative Annotation Systems}


The above-mentioned results yielded implications for the future design of human-AI collaborative annotation systems. Below, we discuss the design implications for using AI-based CAs to best support the co-annotation of online risk data.


\subsection{Reasoning and domain knowledge provided by CAs are valuable resources for co-annotation workflows} We observed that CHAIRA was good at providing reasons for its annotation results. Therefore, the benefit of CHAIRA lies in the interactive nature of the annotation process, which provides partial explainability and is one of the important aspects of human-centered AI-based systems~\cite{vilone2021notions,minh2022explainable}. 
With the most sophisticated prompting approach we had (Few-shot CoT), CHAIRA informed the human coder with a broad knowledge and context of the given text and convinced the human coder to change their mind in multiple instances. 
Therefore, one of the strengths of LLM-based CAs is the ability to provide relevant information (presumably) trained on the entirety of online data, as opposed to human coding which requires extensive training or domain-specific knowledge. This shows the potential strengths and values in human-AI co-annotation, particularly in risk scenarios that require domain-specific knowledge. In addition, the initial reasoning and domain knowledge provided by CAs can be used to inform human coders further on how to design better CA models to support the co-annotation of online risk.

\subsection{Two-way interaction between human coders and CAs is a key to good co-annotation results.} A major benefit of co-coding with CHAIRA was its ability to scale the data annotation with a high degree of inter-rater agreement. To be able to do so, we carefully reviewed the incivility labels where CHAIRA and the human coder disagreed and had two-way conversations with CHAIRA to further understand their reasoning. This two-way interaction in the co-annotation process was useful because, despite access to a large corpus of knowledge, CHAIRA also tended to make some simple mistakes that were quite easy for human coders to spot. For instance, during the interaction with CHAIRA, we realized that CHAIRA could miss the previous conversation and hence, we needed to remind it about our conversation for further annotation tasks, particularly about nuanced incivility (e.g., implicit aspersion). 
Therefore, we added the instruction \textit{``keep implicit incivility in mind,''} in our Two-stage CoT prompt. 
Sometimes, we pointed to the exact incivil expressions in the text that contain implicit aspersion to remind the concept (e.g., Don't you think this expression of ``he’s hoping to stir up the same frenzy and ride that wave?'' could be implicitly incivil?). Then the CHAIRA re-evaluated the text and corrected their answers. 
As such, interactive communication between the human coders and the AI agent is one of the key elements in improving the risk annotations generated by the AI agent.

\subsection{Providing clear examples in carefully designed prompts considering how LLMs process human language is important} 
Providing clear examples of risk cases and descriptions of risk types is crucial when designing human-AI co-annotator models. In our few-shot prompt, we explained that aspersion can be both implicit and explicit, yet CHAIRA failed to recognize implicit aspersion until we guided it through two-way communication. This could be due to confusion caused by slight differences between the risk descriptions and examples provided. For instance, CHAIRA may have interpreted the explicit nature of the ``silly border wall'' in the aspersion example and missed the implicit aspect described. 
Therefore, selecting the right examples and crafting clear descriptions of constructs is critical when working with LLM-based CAs to annotate subjective concepts like online risk. In addition, since CAs generate responses by tokenizing input~\cite{OpenAI_token_2023}, even minor textual changes—such as punctuation—can affect performance. This can limit the CA’s ability to understand text containing abbreviations or spelling variations, which are common in online risk data~\cite{sadeque2019incivility}. Therefore, designing prompts with careful consideration of how LLMs process natural language is essential to building effective collaborative systems for annotating contextualized online risk data.


\subsection{Limitations and Future Directions.} 
In consonance with data ethics, we used de-identified data to develop our annotation tool. We used OpenAI API in the backend as its security policy stipulates that data submitted through the API is not used to train OpenAI models~\cite{OpenAI_security_2024}.
However, future work should also consider building LLM-based CAs with private servers so that the training dataset is not shared via the web. Another limitation of this collaborative annotation is variations in LLM responses. 
Finally, we acknowledge that inductive approaches (e.g., thematic analysis, grounded theory approach), important approaches to building patterns and themes in qualitative work, were not explored in this study. Future work can explore the potential of LLM-based CAs for more inductive analysis that requires an in-depth understanding of the subtleties and complexities of qualitative data. Moving forward, we aim to experiment with our approaches with diverse types of online risk data to gain deeper insights into collaborative annotation between human and LLM-based conversational agent systems. 



\section{Conclusion}
In this study, we built systems to support human-AI collaborative data annotation task and explored the potential benefits and challenges of human-AI collaborative annotation of highly subjective and contextualized online incivility data. The AI missed some implicit risks that human coders easily spotted, conversely, it spotted politically nuanced incivility that human coders overlooked. The design implications and best practices derived from this work can serve as a stepping stone for future research considering similar methods. Our work suggests a path toward combining the relative strengths of humans and AI for scalable data annotation, especially in sensitive or low-resource settings.

\newpage
\appendix
\section{Appendix}

\subsection{Conversation log for two-stage few-shot CoT prompt}
\begin{table}[h]
\caption{Conversation log that was added to the two-stage few-shot CoT prompt}
\small
\begin{tabular}{p{1.8cm}p{10.1cm}}\hline
Data    & All this seems like is puffing and boasting by a politician who is coming up for reelection and has a prospective presidential run in the future. Just trying to get a check mark next to his name for being strong on border and immigration without actually doing much to change, fix, or address the problem at all. All the while crying out at every chance you get how it is your oppositions fault. The real problem here is the consequences of there actions. The U.S citizens whose land is going to be taken from them in imminent domain. Those who are now completely blocked from access to the water way. The years and years of lawsuits these people are going through. \\\hline
AI labeler & Type: Civil. Explanation: The text expresses criticism and frustration towards a politician and their actions, but it does not contain any explicit name-calling, aspersions, lying, vulgarity, pejorative for speech, or other uncivil language. It focuses on the consequences and impact of the politician's actions, which can be seen as a legitimate critique. Therefore, it can be classified as civil.   \\\cline{2-2}
Human labeler  & Some of the incivility, like aspersion, can be implicit and nuanced. What do you think?                                                 \\\cline{2-2}
AI labeler & Type: Incivil. Upon closer examination, the text does contain elements of implicit aspersion directed at the politician. The language used suggests that the politician is simply "puffing and boasting" without actually taking meaningful action to address the border and immigration problem. The text also implies that the politician is blaming their opposition for the issue without taking responsibility themselves. These implicit aspersions contribute to an overall tone of disrespect towards the politician. Therefore, the text can be classified as incivil.                                                                \\\cline{2-2}
Human labeler & Keep implicit incivility in mind, classify the text into "civil" or "incivil."    \\\hline    
\end{tabular}
\label{tab:conversation log}
\end{table}

\newpage
\subsection{Comparison between responses with different prompting approaches}

\begin{table}[h]

\caption{Comparison between the responses for the incivil case generated with zero-shot vs. definition prompting approaches }
\begin{tabular}{p{1.3cm}p{10.6cm}}\hline
Text       & What does combating global climate crisis even mean? Why does anyone thing politicians have answers. They push this agenda for additional tax money. Time magazine said there was a mini ice age coming in the 1970’s. Then Al Gore (internet inventor) said we had global warming that would drown all of Florida and half of New York right now. Now we have climate change. Yes we do, everything changes. Everything has a cycle. Nothing is constant. If the climate didn’t change government would raise taxes telling us they could change the stagnant climate. Wake up people…" \\\hline
Zero-shot response   & This comment is incivil because it resorts to \textbf{personal attacks and insults} toward politicians, insinuates ulterior motives for advocating climate action, and dismisses the scientific consensus on climate change. It also uses \textbf{sarcasm and an accusatory tone} towards the readers, which further contributes to its incivility. \\\cline{2-2}
Definition response & This comment falls under the category of \textbf{``lying''} because it implies that politicians are pushing the climate agenda solely for additional tax money, without providing any evidence or factual basis for this claim. It also contains \textbf{sarcastic remarks and dismissive language, which contributes to the disrespectful tone of the comment.}
\\ \hline
\end{tabular}
\label{tab:uncivil example}
\end{table}

\begin{table}[h]
\caption{Comparison among responses for the implicit incivil case with different prompting approaches}

\begin{tabular}{p{1.4cm}p{10.5cm}}
\toprule
Text          & It’s bad enough that we give weapons to bellicose countries which use those weapons to attack weaker ones, and then spend even more money sending those countries aid. There’s a sick lack of empathy for people right across the border, as you said. So many people here are incapable of putting themselves in another’s shoes, unable to imagine being born into less fortunate circumstances, and having to do whatever it takes to survive and give your children a better chance in a safer country. But then, once they need help in Texas, they want to be bailed out and point the blame at anyone else.  \\ \midrule
Zero-shot response    & This comment expresses a strong opinion and criticism towards the government and the lack of empathy for people in other countries. However, it does so in a respectful and rational manner, addressing a common issue of hypocrisy that some individuals may have. While it may be a controversial statement, \textbf{it does not contain any personal attacks or offensive language, making it a civil comment.}   \\\cline{2-2}
Few-shot  response    & The comment expresses a strong opinion about the topic and may be seen as critical or unkind towards the people who hold a certain viewpoint, but it \textbf{does not contain any explicit name-calling, aspersions, lies, vulgarity, pejorative for speech, or other forms of incivility}. The language used is generally respectful, even though it may convey frustration or disagreement. \\\cline{2-2}
Two-stage CoT response & The text contains elements of incivility. \textbf{It includes implicit aspersions towards certain groups of people who lack empathy and blame others when in need.} The language used, such as "incapable," "unable," and "point the blame," conveys a disrespectful tone towards these individuals. Additionally, the text exhibits a divisive attitude by highlighting the perceived hypocrisy of those in Texas. Therefore, the text can be classified as incivil.  \\ \bottomrule
\end{tabular}
\label{Example responses}
\end{table}

\newpage
\bibliographystyle{splncs04}
\bibliography{00-References}

\begin{thebibliography}{10}
\providecommand{\url}[1]{\texttt{#1}}
\providecommand{\urlprefix}{URL }
\providecommand{\doi}[1]{https://doi.org/#1}

\bibitem{amin2023will}
Amin, M.M., Cambria, E., Schuller, B.W.: Will affective computing emerge from
  foundation models and general ai? a first evaluation on chatgpt. arXiv
  preprint arXiv:2303.03186  (2023)

\bibitem{brown2020language}
Brown, T., Mann, B., Ryder, N., Subbiah, M., Kaplan, J.D., Dhariwal, P.,
  Neelakantan, A., Shyam, P., Sastry, G., Askell, A., et~al.: Language models
  are few-shot learners. Advances in neural information processing systems
  \textbf{33},  1877--1901 (2020)

\bibitem{clay2003context}
Clay-Warner, J.: The context of sexual violence: Situational predictors of
  self-protective actions. Violence and victims  \textbf{18}(5),  543--556
  (2003)

\bibitem{coe2014online}
Coe, K., Kenski, K., Rains, S.A.: Online and uncivil? patterns and determinants
  of incivility in newspaper website comments. Journal of Communication
  \textbf{64}(4),  658--679 (2014)

\bibitem{esau2022content}
Esau, K.: Content analysis in the research field of incivility and hate speech
  in online communication. In: Standardisierte Inhaltsanalyse in der
  Kommunikationswissenschaft--Standardized Content Analysis in Communication
  Research: Ein Handbuch-A Handbook, pp. 451--461. Springer Fachmedien
  Wiesbaden Wiesbaden (2022)

\bibitem{gervais2015incivility}
Gervais, B.T.: Incivility online: Affective and behavioral reactions to uncivil
  political posts in a web-based experiment. Journal of Information Technology
  \& Politics  \textbf{12}(2),  167--185 (2015)

\bibitem{gilardi2023chatgpt}
Gilardi, F., Alizadeh, M., Kubli, M.: Chatgpt outperforms crowd workers for
  text-annotation tasks. Proceedings of the National Academy of Sciences
  \textbf{120}(30),  e2305016120 (2023)

\bibitem{han2018civility}
Han, S.H., Brazeal, L.M., Pennington, N.: Is civility contagious? examining the
  impact of modeling in online political discussions. Social Media+ Society
  \textbf{4}(3),  2056305118793404 (2018)

\bibitem{hosseinmardi2015analyzing}
Hosseinmardi, H., Mattson, S.A., Rafiq, R.I., Han, R., Lv, Q., Mishra, S.:
  Analyzing labeled cyberbullying incidents on the instagram social network.
  In: International conference on social informatics. pp. 49--66. Springer
  (2015)

\bibitem{huang2023chatgpt}
Huang, F., Kwak, H., An, J.: Is chatgpt better than human annotators? potential
  and limitations of chatgpt in explaining implicit hate speech. arXiv preprint
  arXiv:2302.07736  (2023)

\bibitem{jhaver2018online}
Jhaver, S., Ghoshal, S., Bruckman, A., Gilbert, E.: Online harassment and
  content moderation: The case of blocklists. ACM Transactions on
  Computer-Human Interaction (TOCHI)  \textbf{25}(2),  1--33 (2018)

\bibitem{jiang2021supporting}
Jiang, J.A., Wade, K., Fiesler, C., Brubaker, J.R.: Supporting serendipity:
  Opportunities and challenges for human-ai collaboration in qualitative
  analysis. Proceedings of the ACM on Human-Computer Interaction
  \textbf{5}(CSCW1),  1--23 (2021)

\bibitem{kim2022stylette}
Kim, T.S., Choi, D., Choi, Y., Kim, J.: Stylette: Styling the web with natural
  language. In: Proceedings of the 2022 CHI Conference on Human Factors in
  Computing Systems. pp. 1--17 (2022)

\bibitem{kojima2022large}
Kojima, T., Gu, S.S., Reid, M., Matsuo, Y., Iwasawa, Y.: Large language models
  are zero-shot reasoners. Advances in neural information processing systems
  \textbf{35},  22199--22213 (2022)

\bibitem{kuzman2023chatgpt}
Kuzman, T., Mozetic, I., Ljube{\v{s}}ic, N.: Chatgpt: Beginning of an end of
  manual linguistic data annotation? use case of automatic genre
  identification. ArXiv, abs/2303.03953  (2023)

\bibitem{lai2022human}
Lai, V., Carton, S., Bhatnagar, R., Liao, Q.V., Zhang, Y., Tan, C.: Human-ai
  collaboration via conditional delegation: A case study of content moderation.
  In: Proceedings of the 2022 CHI Conference on Human Factors in Computing
  Systems. pp. 1--18 (2022)

\bibitem{liu2023summary}
Liu, Y., Han, T., Ma, S., Zhang, J., Yang, Y., Tian, J., He, H., Li, A., He,
  M., Liu, Z., et~al.: Summary of chatgpt-related research and perspective
  towards the future of large language models. Meta-Radiology p. 100017 (2023)

\bibitem{lu2024differences}
Lu, Q., Peng, X.: Differences in knowledge adoption among task types in
  human-ai collaboration under the chronic disease prevention scenario. In:
  International Conference on Information. pp. 213--231. Springer (2024)

\bibitem{mackeprang2019discovering}
Mackeprang, M., M{\"u}ller-Birn, C., Stauss, M.T.: Discovering the sweet spot
  of human-computer configurations: A case study in information extraction.
  Proceedings of the ACM on Human-Computer Interaction  \textbf{3}(CSCW),
  1--30 (2019)

\bibitem{matias2019preventing}
Matias, J.N.: Preventing harassment and increasing group participation through
  social norms in 2,190 online science discussions. Proceedings of the National
  Academy of Sciences  \textbf{116}(20),  9785--9789 (2019)

\bibitem{mcdonald2019reliability}
McDonald, N., Schoenebeck, S., Forte, A.: Reliability and inter-rater
  reliability in qualitative research: Norms and guidelines for cscw and hci
  practice. Proceedings of the ACM on human-computer interaction
  \textbf{3}(CSCW),  1--23 (2019)

\bibitem{minh2022explainable}
Minh, D., Wang, H.X., Li, Y.F., Nguyen, T.N.: Explainable artificial
  intelligence: a comprehensive review. Artificial Intelligence Review pp.
  1--66 (2022)

\bibitem{OpenAI2023}
OpenAI: Introducing chatgpt (2023), \url{https://openai.com/blog/chatgpt}

\bibitem{OpenAI_security_2024}
OpenAI: Security \& privacy (2024), \url{https://openai.com/security}

\bibitem{OpenAI_token_2023}
OpenAI: What are tokens and how to count them? (2024),
  \url{https://help.openai.com/en/articles/4936856-what-are-tokens-and-how-to-count-them}

\bibitem{oz2018twitter}
Oz, M., Zheng, P., Chen, G.M.: Twitter versus facebook: Comparing incivility,
  impoliteness, and deliberative attributes. New media \& society
  \textbf{20}(9),  3400--3419 (2018)

\bibitem{park2023misinformation}
Park, J., Ellezhuthil, R.D., Isaac, J., Mergerson, C., Feldman, L., Singh, V.:
  Misinformation detection algorithms and fairness across political ideologies:
  The impact of article level labeling. In: Proceedings of the 15th ACM Web
  Science Conference 2023. pp. 107--116 (2023)

\bibitem{park2023towards}
Park, J., Gracie, J., Alsoubai, A., Stringhini, G., Singh, V., Wisniewski, P.:
  Towards automated detection of risky images shared by youth on social media.
  In: Companion Proceedings of the ACM Web Conference 2023. pp. 1348--1357
  (2023)

\bibitem{park2022background}
Park, J., Singh, V.K.: How background images impact online incivility.
  Proceedings of the ACM on Human-Computer Interaction  \textbf{6}(CSCW2),
  1--23 (2022)

\bibitem{rains2017incivility}
Rains, S.A., Kenski, K., Coe, K., Harwood, J.: Incivility and political
  identity on the internet: Intergroup factors as predictors of incivility in
  discussions of news online. Journal of Computer-Mediated Communication
  \textbf{22}(4),  163--178 (2017)

\bibitem{rheu2021systematic}
Rheu, M., Shin, J.Y., Peng, W., Huh-Yoo, J.: Systematic review: Trust-building
  factors and implications for conversational agent design. International
  Journal of Human--Computer Interaction  \textbf{37}(1),  81--96 (2021)

\bibitem{sadeque2019incivility}
Sadeque, F., Rains, S., Shmargad, Y., Kenski, K., Coe, K., Bethard, S.:
  Incivility detection in online comments. In: Proceedings of the eighth joint
  conference on lexical and computational semantics (* SEM 2019). pp. 283--291
  (2019)

\bibitem{singh2017toward}
Singh, V.K., Ghosh, S., Jose, C.: Toward multimodal cyberbullying detection.
  In: Proceedings of the 2017 CHI Conference Extended Abstracts on Human
  Factors in Computing Systems. pp. 2090--2099 (2017)

\bibitem{song2024preference}
Song, F., Yu, B., Li, M., Yu, H., Huang, F., Li, Y., Wang, H.: Preference
  ranking optimization for human alignment. In: Proceedings of the AAAI
  Conference on Artificial Intelligence. vol.~38, pp. 18990--18998 (2024)

\bibitem{tamura2024influence}
Tamura, T., Ito, H., Oyama, S., Morishima, A.: Influence of ai’s uncertainty
  in the dawid-skene aggregation for human-ai crowdsourcing. In: International
  Conference on Information. pp. 232--247. Springer (2024)

\bibitem{tang2024pdfchatannotator}
Tang, Y., Chang, C.M., Yang, X.: Pdfchatannotator: A human-llm collaborative
  multi-modal data annotation tool for pdf-format catalogs. In: Proceedings of
  the 29th International Conference on Intelligent User Interfaces. pp.
  419--430 (2024)

\bibitem{tinsley2000interrater}
Tinsley, H.E., Weiss, D.J.: Interrater reliability and agreement. In: Handbook
  of applied multivariate statistics and mathematical modeling, pp. 95--124.
  Elsevier (2000)

\bibitem{vilone2021notions}
Vilone, G., Longo, L.: Notions of explainability and evaluation approaches for
  explainable artificial intelligence. Information Fusion  \textbf{76},
  89--106 (2021)

\bibitem{wei2022chain}
Wei, J., Wang, X., Schuurmans, D., Bosma, M., Xia, F., Chi, E., Le, Q.V., Zhou,
  D., et~al.: Chain-of-thought prompting elicits reasoning in large language
  models. Advances in Neural Information Processing Systems  \textbf{35},
  24824--24837 (2022)

\bibitem{zhang2022would}
Zhang, B., Ding, D., Jing, L.: How would stance detection techniques evolve
  after the launch of chatgpt? arXiv preprint arXiv:2212.14548  (2022)

\bibitem{zhang2023qualigpt}
Zhang, H., Wu, C., Xie, J., Kim, C., Carroll, J.M.: Qualigpt: Gpt as an
  easy-to-use tool for qualitative coding. arXiv preprint arXiv:2310.07061
  (2023)

\bibitem{zhang2022automatic}
Zhang, Z., Zhang, A., Li, M., Smola, A.: Automatic chain of thought prompting
  in large language models. arXiv preprint arXiv:2210.03493  (2022)

\bibitem{zhou2022large}
Zhou, Y., Muresanu, A.I., Han, Z., Paster, K., Pitis, S., Chan, H., Ba, J.:
  Large language models are human-level prompt engineers. arXiv preprint
  arXiv:2211.01910  (2022)

\end{thebibliography}




\end{document}